\documentclass{article}

\usepackage{amsmath,amssymb,amsfonts,amsthm}
\usepackage[caption=false]{subfig}
\usepackage{microtype}
\usepackage{hyperref}
\usepackage{braket}
\usepackage{graphicx}

\usepackage{authblk}

\newcommand{\Z}{\mathbb{Z}}
\newcommand{\R}{\mathbb{R}}

\newcommand{\eqdef}{\overset{\underset{\mathrm{def}}{}}{=}}
\newcommand{\norm}[1]{ \left\| #1 \right\| }

\newcommand{\DS}{D_{S^2}}
\newcommand{\DB}{D_{S^2} + c B}
\newcommand{\fdspace}{\mathcal{D}}

\DeclareMathOperator{\tr}{tr}

\theoremstyle{plain}

\theoremstyle{remark}

\theoremstyle{definition}

\newcommand{\constraint}{\delta}
\newcommand{\cstar}{C$^*$ }
\newcommand{\spinc}{spin$^\mathbf{C}$ }
\newcommand{\mc}[1]{\mathcal{#1}}

\newlength{\maxfigwidth}
\newlength{\maxfigheight}

\makeatletter
\def\blfootnote{\xdef\@thefnmark{}\@footnotetext}
\makeatother

\begin{document}

\setlength{\maxfigwidth}{1.0\textwidth}

\setlength{\maxfigheight}{0.4\textheight}

\title{Understanding truncated non-commutative geometries through computer simulations} 



\author[1]{L. Glaser}
\affil[1]{University of Vienna, Vienna, Austria}

\author[2]{A.B. Stern}
\affil[2]{Institute of Mathematics, Astrophysics and Particle Physics, Radboud University, Nijmegen, the Netherlands}

\date{February 15th, 2020}

\maketitle 

\begin{abstract}
When aiming to apply mathematical results of non-commutative geometry to physical problems the question arises how they translate to a context in which only a part of the spectrum is known.

In this article we aim to detect when a finite-dimensional triple is the truncation of the Dirac spectral triple of a spin manifold.
To that end, we numerically investigate the restriction that the higher Heisenberg equation [A. H.
Chamseddine, A. Connes, and V. Mukhanov, Journal of High Energy Physics, 98
(2014)] places on a truncated Dirac operator.
We find a bounded perturbation of the Dirac operator on the Riemann sphere that induces the same Chern class.
\end{abstract}


\blfootnote{This article may be downloaded for personal use only. Any other use requires prior permission of the author and AIP Publishing. This article appeared in Journal of Mathematical Physics \textbf{61}, 033507 (2020) and may be found at
  \url{https://doi.org/10.1063/1.5131864}.}

\section{Introduction}
The spectral viewpoint is central to non-commutative geometry, which makes it a natural framework to investigate the relation between energy and geometry.
To understand low-energy (that is, physical) observations, we need to be able to distinguish commutative spectral triples from classically meaningless configurations, using only low-energy data.

Connes' spectral reconstruction theorem~\cite{Connes:SpectralCharacterizationManifolds} tells us when a spectral triple $(A, H, D)$ is the Dirac triple of a \spinc manifold.
However, checking the conditions under which the theorem holds requires knowledge of all spectral information: they can not be applied when we only consider a finite part of the frequency (energy) spectrum.
That is, the usual spectral expressions do not reveal much about the nature of the universe to an observer with access to only finite spectral information.

This is highly relevant when applying non-commutative geometry to physical problems, since in realistic systems only approximate knowledge is available.
It is also highly relevant when using spectral triples to discretize geometries through finite algebras and Hilbert spaces, and in most attempts to use numerical methods to explore spectral triples.

In order to engage the issue, we explore whether it is possible to use the higher Heisenberg equation~\cite{C.C.M:GeometryQuantumBasics} to detect, at a finite frequency level, whether a given truncated spectral triple corresponds to a \spinc manifold.
The analysis starts with a computer simulation of the higher Heisenberg constraint (introduced below) on the sphere, which leads to a new analytic solution of the corresponding equation.
Lastly, the methods from the companion paper \cite{GlaserSternDistance19} are applied to generate and visualise finite metric graphs that represent (what is argued to be) the metric space corresponding to the finite-scale geometries involved.

The remainder of this introduction is structured as follows.
Section \ref{sec:backgroundncg} briefly introduces the relevant concepts from noncommutative geometry, such as spectral triples, the spectral action principle, and the relation of the latter to observations based on finite spectra.
Section \ref{sec:intropartialspectra} expands on the notion of information contained in finite spectra and introduces the problem of detecting `commutativity' at finite scale, whereafter Section \ref{sec:higherheisenberg} introduces the higher Heisenberg relation as a possible approach to that problem and gives a brief overview of the structure of the paper itself.

\subsection{Background: noncommutative geometry and the cutoff scale}
\label{sec:backgroundncg}
By Gelfand duality, a (compact Hausdorff) space $X$ may be entirely understood in terms of the algebra $C(X)$ of continuous functions. Moreover, each commutative unital $C^*$-algebra is of this form $C(X)$ for some $X$.

Noncommutative geometry starts by the observation that we can extend this duality to \spinc manifolds: the \spinc manifold $M$ (and, therefore, its metric) can be described uniquely in terms of the \emph{spectral triple} $(C^\infty(M), H, D)$, where $D$ is the associated Dirac operator and $H$ is a Hilbert space of spinors.

This description of spin geometry in terms of operators on Hilbert spaces then allows one to extend many spin-geometric notions to the study of more general geometric objects, the noncommutative spectral triples\footnote{Here, $A$ is a possibly noncommutative C$^*$-algebra, corresponding to the `topological' aspect of the noncommutatige geometry, and $D$ a possibly unbounded selfadjoint operator, corresponding to the `metric' aspect thereof, both represented on a Hilbert space $H$. See \cite{gracia2013elements} for an introduction.} $(A, H, D)$.
In particular, the resulting flexibility allows one to use the same language to describe both ordinary spin geometry and the field theories common in particle physics~\cite{Chamseddine:2006ep}. A very simple choice of algebra, together with the \emph{spectral action principle}~\cite{Chamseddine:1996zu} (see below) leads to the standard model, minimally coupled to general relativity.

One interesting feature of this latter formulation is that while the classical (metric) geometry is described through an infinite-dimensional algebra and Hilbert space, the particles of the standard model are encoded in a finite-dimensional non-commutative algebra.
Fundamentally finite dimensional spectral triples allow for a description of spaces that are discretized, but still retain their original symmetry group.
Examples of these, often called fuzzy, spaces are the fuzzy sphere~\cite{Grosse1995}, fuzzy projective spaces~\cite{BALACHANDRAN2002184} or the fuzzy torus~\cite{Dolan_2003}.
General finite spectral triples have been classified~\cite{krajewski1998classification,cacic2011moduli, chamseddine2008standard} and parametrized~\cite{Barrett:2015naa}.
In the present paper, however, we will be concerned with \emph{truncated}, not fundamentally finite, spectral triples. Following~\cite{ConnesSuij}, these truncated spectral triples should perhaps properly be called \emph{operator system spectral triples}.

By the assumption of diffeomorphism invariance, all observables in pure gravity -- including the action -- must be expressible in terms of global geometric invariants.
The spectral action principle \cite{chamseddine_spectral_1997} in noncommutative geometry asserts that, moreover, the action should be formulated in terms of the spectrum of the Dirac operator $D$ alone.
The identification of such global invariants with zeta residues allows them to be written in terms of asymptotic traces of $D$, and this induces the prescription
\[
   S(D) = \tr(f(D / \Lambda))
\]
for the bare action, where $f$ should be a suitable smooth cutoff function.
The scale parameter $\Lambda$ controls the relative contributions of Dirac eigenvalues.
At finite cutoff scale $\Lambda$ we are then automatically invited to think of the corresponding system as described by a finite-rank, \emph{truncated} Dirac operator.

Recent work has started numerically exploring the path integral,
\begin{align} \label{eq:Z}
\mc{Z} = \int \mathrm{d}D e^{- \mc{S}(D)},
\end{align}
with $\mc{S}$ a trace of powers of $D$, over finite-rank Dirac operators, as a possible nonperturbative description for quantum gravitational phenomena~\cite{Barrett_Glaser_2016,glaserScalingBehaviourRandom2017,Barrett:2019aig}.

\subsection{Geometry at finite scale}
\label{sec:intropartialspectra}
Spectral descriptions of continuum geometry involve infinite-dimensional algebras and Hilbert spaces. If these are to be applied to physics involving measurement at finite energies, to be captured in computer simulations or to be described approximately, we must understand \emph{how (much) information can be contained in partial spectra}.
This involves extending the tools that have been developed to understand infinite-dimensional non-commutative geometries to truncated spectral triples, as has been done for the residue functionals in~\cite{Stern_2018}.

A particular difficulty, which is central to the present paper, relates to the recognition of (possibly almost-commutative) manifolds at the truncated level. In carrying out the path integral \eqref{eq:Z}, for instance, one should in principle restrict to Dirac operators that actually correspond to (possibly almost-commutative) \spinc structures for the given (fixed) manifold $M$, just like the path integral in Euclidean quantum gravity restricts the integration to fields that describe Riemannian metrics as opposed to being fully arbitrary. However, it is a priori unclear what this restriction means for the integration variable $D$.

Although Connes' reconstruction theorem \cite{Connes:SpectralCharacterizationManifolds} allows us to detect when a spectral triple $(A, H, D)$ corresponds to the Dirac triple on a spin manifold, it is not clear how to implement those conditions as a constraint on an integral over operators $D$.
Moreover, it is not clear when a \emph{finite-rank} Dirac operator $D$ corresponds to a cutoff of such a spin geometry.
This complicates the proposed identification of path integrals over finite-rank Dirac operators with finite-scale path integrals over spin geometries.
The one-sided higher Heisenberg equation recalled below (and more generally, its two-sided version) offers a possible approach to constraining the domain of integration in \eqref{eq:Z}.

\subsection{The higher Heisenberg equation }
\label{sec:higherheisenberg}
In \cite{C.C.M:GeometryQuantumBasics} Chamseddine, Connes and Mukhanov introduce a non-commutative analogue to the Heisenberg relation of quantum mechanics. This `higher Heisenberg equation' neatly captures the relation between the scalar fields (smooth functions) and the Dirac operator that is central to noncommutative geometry in a single \emph{algebraic} equation.
The one-sided version of this equation, applicable to (disjoint sums of) even-dimensional $n$-spheres, works as follows.
There trivially exists a covering $\phi: M \to S^n$; let its components be denoted by $Y^i$, $1 \leq i \leq n$.
Then, the section $Y = Y^i \Gamma_i$ of the trivial Clifford bundle of rank $2^{n/2}$ satisfies $Y^2 = 1$ and $Y^* = Y$, and moreover the Chern character of the idempotent $e = \frac12(1 + Y)$ satisfies
\begin{align}\label{eq:CCM}
  \frac{1}{n!} \langle Y \underbrace{[Y,D] \dots [Y,D]}_{\text{repeated $n$ times}} \rangle = \gamma,
\end{align}
where $\gamma$ is the grading on the spinor bundle and $\langle \cdot \rangle$ denotes the $C^\infty(M)$-valued fiberwise trace on the Clifford algebra bundle.
If a \emph{general} Riemannian manifold $M$ admits such $Y$, moreover, they must necessarily be of the form considered above, ensuring that $M$ is a disjoint sum of even-dimensional spheres.
The $Y_i$ then generate $C^\infty(S^n)$ and the spectral triple $(C^\infty(M), H, D)$ is unitarily equivalent to the direct sum of any splitting of $(C^\infty(S^n), H, D)$ into irreducible components.

For more general (spin) $M$, the real structure on the spinor bundle induces a two-sided version of the equation above, corresponding to a map $\phi \times \phi'$ that induces a (not necessarily isometric) embedding $M \to S^n \times S^n$. We are presently concerned only with the one-sided equation as a first example.

In this article we propose to use the higher Heisenberg relation to constrain general selfadjoint matrices $D$, in order to induce them to correspond to truncated Dirac operators of reasonable Riemannian geometries on the underlying manifold.
Computer simulations then allow us to explore numerically the effects of this constraint.

A real spectral triple consists of $(\mc{A},\mc{H},D)$ together with a real structure $J$ and a chirality $\gamma$ that satisfy a number of conditions.
An introduction can be found e.g. in~\cite{gracia2013elements}.
One axiom that has special significance, is the first order condition
\[
  \left[ [D,a],J b^* J^{-1} \right] = 0 \qquad \forall \,  a,b \in \mc{A},
\]
which ensures that $D$ acts as first-order differential operator in the commutative case, and is the second algebraic constraint (besides the one corresponding to the higher Heisenberg equation) appearing in Connes' reconstruction theorem.

To recover the metric on a \spinc manifold from the corresponding spectral triple, one can define a metric on the space of states $\omega_1, \omega_2 \in S(\mc{A})$,
\begin{align}\label{eq:distance}
  d(\omega_1,\omega_2)=\sup{a\in \mc{A}} \{ \omega_1(a)-\omega_2(a) | ||[D,a]|| \leq 1\}.
\end{align}
In the commutative case, the pure states correspond to atomic measures, that is, points, on the underlying manifold.
In the companion paper~\cite{GlaserSternDistance19} we use this definition of distance, together with a notion of locality, to associate \emph{finite} metric spaces to truncated non-commutative geometries.

In section \ref{sec:decibel} we explain the truncation and our simulations methods and present results for the circle and the two-sphere.
This section in particular discusses the reasoning behind our choice of truncation, how it is implemented and some possible problems in this choice.
In section \ref{sec:analytic} we show that one of the Dirac operators found in the previous section is a better solution to the Heisenberg relation, while not strictly belonging to a spectral triple in the infinite size limit.
In our conclusion, section \ref{sec:conclusion}, we summarize the results and collect some questions that are opened by our work.

\section{The Heisenberg relation in simulations}
\label{sec:decibel}

\label{discussing_truncation}
In noncommutative geometry one describes a spin manifold in terms of the associated spectral triple $(A, H, D)$.
From a mathematical perspective, it is desirable to be able to describe such a spectral triple as a limit of finite-dimensional data of increasing precision, just like one can describe a Riemannian manifold as a Gromov-Hausdorff limit of finite metric spaces.
From a physical perspective, the same desire results from the view that one should be able to gain at least \emph{some} information about the geometry by probing it at finite energies.

One natural approach to such a `cutoff' of the geometric data $(A, H, D)$ is to pick a scale $\Lambda$, then define
\[
  P_\Lambda \eqdef \chi_{[-\Lambda,\Lambda]}(D)
\]
to be the spectral projection onto the eigenspaces of $D$ of eigenvalue $|\lambda| \leq \Lambda$, and then take the finite-dimensional data
\begin{align}
  (P_\Lambda A P_\Lambda, P_\Lambda H, P_\Lambda D )
\end{align}
as our starting point.
This point of view is further explored in as of yet unpublished work by Connes and van Suijlekom~\cite{ConnesSuij}.
In this setting, the companion paper~\cite{GlaserSternDistance19} reconstructs (asymptotically) spin manifolds $M$ from the data $(P_\Lambda C^\infty(M) P_\Lambda, P_\Lambda H, P_\Lambda D_M)$ associated to the commutative spectral triple $(C^\infty(M), H, D_M)$.
Some properties of the induced metric on the state spaces of $P_\Lambda  A P_\Lambda$ and $A$ were previously investigated in~\cite{d2014spectral}, and for the sphere specifically in~\cite{d2013metric}. 

\subsubsection{The truncated higher Heisenberg equation}
All spin manifolds of dimension $\leq 4$ satisfy (the two-sided version of) the higher Heisenberg equation \eqref{eq:CCM}, whereas clearly not all spectral triples do.
This suggests to use the equation to recognize many cases in which a spectral triple does \emph{not} correspond to a spin manifold, without needing to check the rather elusive conditions of the spectral reconstruction theorem.
We will extend this tool to the finite-dimensional data $(P_\Lambda A P_\Lambda, P_\Lambda H, P_\Lambda D)$ introduced above, and explore what type of truncated triple solves the truncated higher Heisenberg relation.

Given a solution $Y, D$ of equation \eqref{eq:CCM} and the spectral projection $P_\Lambda = \chi_{[-\Lambda,\Lambda]}(D)$, the defect
\begin{equation}\label{eq:con1}
  \constraint(Y_\Lambda,D_\Lambda,\gamma_\Lambda) \eqdef \langle Y_\Lambda [D_\Lambda,Y_\Lambda]^n \rangle - n! k \gamma_\Lambda
\end{equation}
strongly converges (superpolynomially) to zero as $\Lambda \to \infty$.
Simple examples like the circle (see below) show, however, that we cannot expect the defect to converge to zero in any Schatten $p$-norm including $p=\infty$.
One wonders then how strongly requiring equation \eqref{eq:con1} with the finite $Y_\Lambda, \gamma_\Lambda$ restricts the spectral triple.

The direct approach to this question starts by searching for an operator $D'$ on $P_\Lambda H$ that comes at least close to solving \eqref{eq:CCM} in the sense of minimizing the constraint
\begin{equation} \label{eq:CCMconN}
  \norm{\constraint(Y_\Lambda,D',\gamma_\Lambda)}_2^2 = \norm{\langle Y_\Lambda [D'_\Lambda,Y_\Lambda]^n \rangle - n! k \gamma_\Lambda}^2_2,
\end{equation}
for fixed $\Lambda$.
The Hilbert-Schmidt norm is a natural choice here; all Schatten norms are equivalent in finite dimensions and this is the least computationally expensive among them.
This, then, is the constraint whose solutions we investigate numerically below:
\begin{itemize}
\item Fix a cutoff $\Lambda$,
\item Take $P_\Lambda, Y_\Lambda,\gamma_\Lambda$ from the corresponding commutative spectral triple (that is, here, from the circle and the (spin) sphere),
\item Look for the arguments $D'_\Lambda$ (matrices of dimension $\operatorname{rank} P_\Lambda$) that minimize \eqref{eq:CCMconN}.
\end{itemize}
The second step means that, for the sphere, the possible matrix size of the truncations will be restricted to the sums of multiplicities of eigenspaces.
To have some more freedom in the choice of matrix size for $D'_\Lambda$ one could, instead of $P_\Lambda$, use some other projection in its commutant.
It seems, however, that in the cases of the circle and the sphere doing so would introduce a further defect in $\constraint(Y_\Lambda,D_{S^n,\Lambda},\gamma_\Lambda)$.

\subsection{Computation}
In order to numerically investigate the behaviour of \eqref{eq:CCMconN} in practice, we use an annealing type algorithm.
Simulated annealing algorithms find optima of a given function by running a random walk in its domain, with transition probability depending on the value of the optimized function and a global `temperature' parameter $T$ that is decreased in time.
The algorithm we use is called thermal annealing, and controls the temperature by postulating that the information theoretic and thermodynamic entropy densities must agree~\cite{devicentePlacementThermodynamicSimulated2003}.
This is a convenient choice for our problem since it has few free parameters, and we are only interested in the final result.
The free parameters in question are a constant $c$ which governs the speed at which the temperature is lowered and the final temperature $T_f$. Any choice of $c$ that does not lead to freezing out of the system before equilibrium is reached is valid, while the final temperature governs how strongly the system is allowed to fluctuate around the final state.
We set $T_f= 0.001$ and adjust $c$ to the simulations in question, testing several $c$ to ensure the results are equivalent.

The annealing algorithm runs until some $T<T_f$ is reached\footnote{The nature of the algorithm means that we do not have perfect control of the finite temperature, however the exact finite temperature is not important in our case.}, and then simulate the system at this low temperature for a while.
The quantities of interest to us are then the configuration with the lowest value of the constraint, as well as an average over the states at the final temperature.

\subsection{The circle as a simple example}
A first example of an algebraic relation, analogous to \eqref{eq:CCM}, whose solution describes a spin manifold is as follows \cite{Connes_2000}.
Assume that $U \in B(H)$ is unitary and $D$ is a selfadjoint unbounded operator on $H$ such that $0 \in \sigma(D)$ and $D^{-1} \in L^{(1,\infty)}(H)$.
Assume, moreover, that the pair $U, D$ is represented irreducibly.
Then, if $U$ and $D$ satisfy
\begin{align}\label{eq:circle}
U^*[D,U]=1\;,
\end{align}
the triple $(A,H,D)$, where $A$ is a dense subalgebra of the \cstar algebra generated by $U$, is unitarily isomorphic to the spectral triple $(C^\infty(S^1), L^2(S^1), D_{S^1})$ that describes the circle.
Under such an isomorphism $U$ is mapped to the generator $\theta \mapsto e^{i \theta}$ of $C(S^1)$ (up to the obvious phase ambiguity in equation \eqref{eq:circle}).

Given the spectral projection $P_\Lambda$ as in section \ref{discussing_truncation}, the operator $U_\Lambda = P_\Lambda U P_\Lambda$ is no longer unitary and even nilpotent, so \eqref{eq:circle}, with $U$ replaced by $U_\Lambda$, cannot be solved in $D$.

The corresponding version of \eqref{eq:CCMconN} is
\begin{align}\label{eq:CCM_circle}
  \| \constraint(U_\Lambda, D_\Lambda) \|^2_2 = || U_\Lambda [D_\Lambda,U_\Lambda]-1||^2_2 \;.
\end{align}
In order to counter the spurious symmetry $D \mapsto D + c I$ of \eqref{eq:circle}, we demand that $D_\Lambda$ additionally satisfies $DJ = JD$, where $J$ is the real structure corresponding to the pointwise complex conjugation map on $L^2(S^1)$.
This ensures that the spectrum of $D_\Lambda$ is symmetric around $0$ and is implemented as $D_\Lambda \eqdef J^* H_\Lambda J$, where $-i H_\Lambda \in B(P_\Lambda H)$ is real antisymmetric.

Using the constraint~\eqref{eq:CCM_circle} as a weight for thermal annealing we collect two types of observations.
On the one hand, we measure the Dirac operator that leads to the smallest value of the constraint.
This is ideally going to be very close to the Dirac operator for the circle.
To compensate for small numerical fluctuations, we also measure $500$ times after the low final temperature is reached and average these measurements.

\begin{figure}
  \centering
  \includegraphics[width=\maxfigwidth]{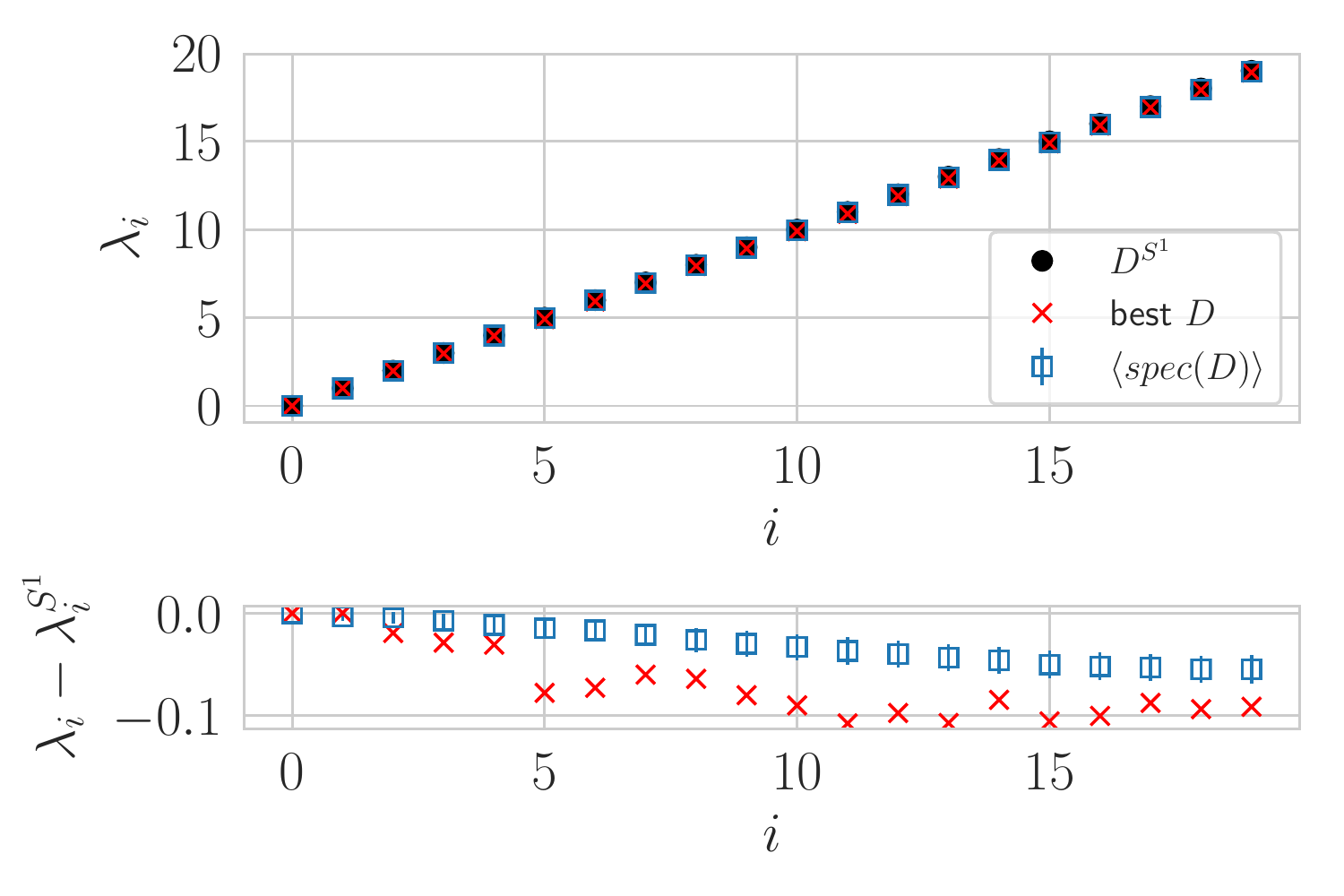}
  \caption{Comparing the eigenvalues of the Dirac operator with the smallest value of the constraint to that of the average over operators (with error indicating the statistical fluctutations) and the exact circle.
  The results are all so close together that we can not distinguish them in the upper plot, the lower plot shows only the difference between the simulation results and the exact numbers.}
  \label{fig:circle_ev}
\end{figure}

In Figure \ref{fig:circle_ev} we see that the eigenvalues of the simulated Dirac operators turn out very close to those of the circle Dirac.
They can not be distinguished in the upper plot, while the lower plot shows the difference from the analytic spectrum for the average and the best eigenvalues.
The small difference is an effect of the cutoff, which is also reinforced by the difference being larger for larger eigenvalues.

\begin{figure}
\centering
{\includegraphics[width=\maxfigwidth]{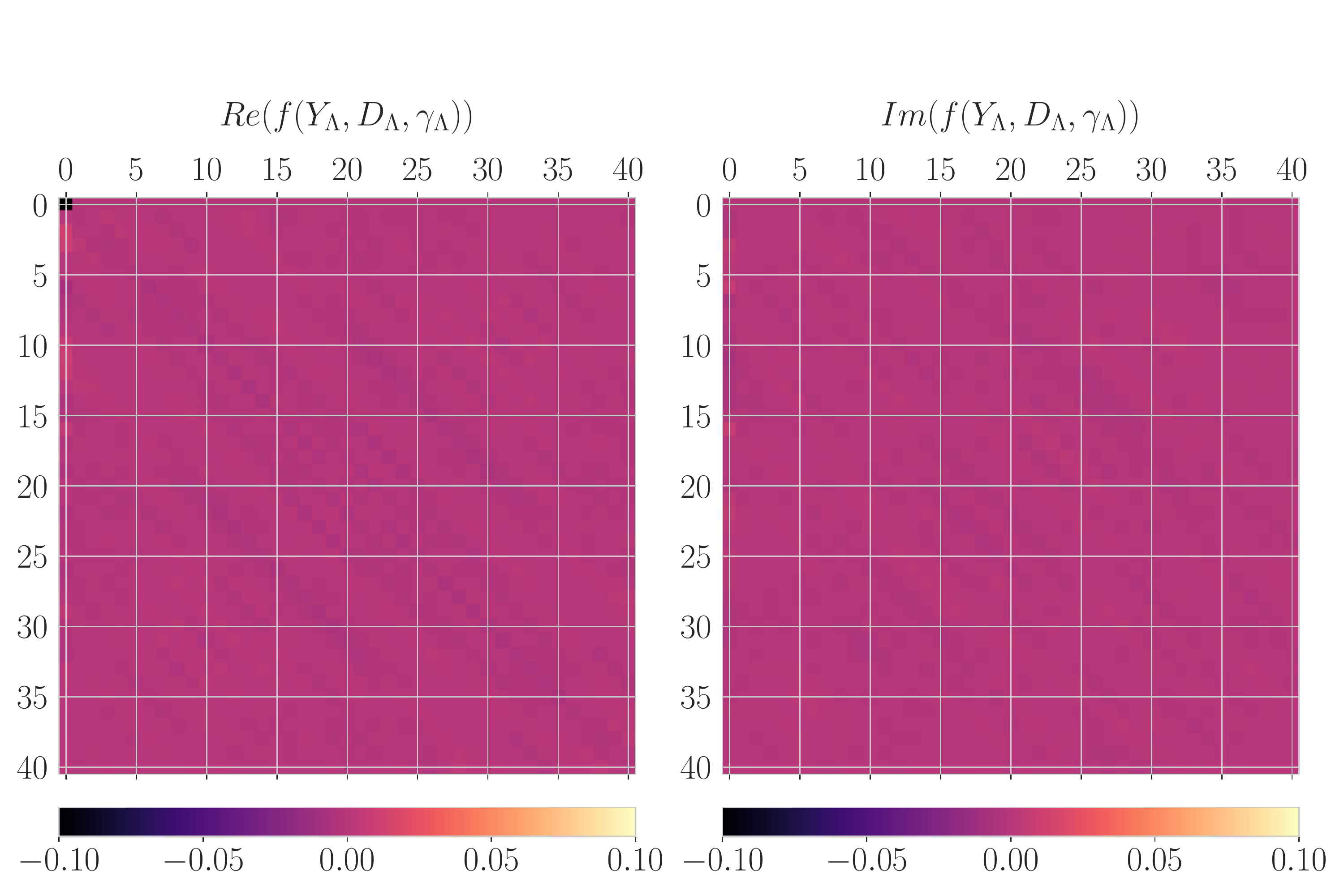}}
\caption{Heatmap plot of the matrix Heisenberg relation averaged over Dirac operators. Each square in the plot corresponds to one matrix element, with the color of the square indicating the value of the element.
This plot shows clearly that only the upper left corner element deviates from $0$ considerably, as does that of $\constraint(U_\Lambda,D_{S^1,\Lambda})$.}
 \label{fig:CCM_circle}
\end{figure}

Another way to visualize the results is to plot the matrix entries $\constraint(U_\Lambda,D_\Lambda)_{ij}$ of the constraint in a heat map.
This makes it possible to not only see how large the violations of the equation are, but also to identify patterns in the defect.
We show this in Figure~\ref{fig:CCM_circle}. While there is some small deviation from $0$ along the diagonal, the main deviation is concentrated in the uppermost left entry, corresponding to the kernel of $U^*$.
This entry is of value $\sim -1$.
Since the defect $U_\Lambda^*[D_{S^1,\Lambda},U_\Lambda]-P_\Lambda$ equals the projection onto that kernel, it is not surprising to find the maximum there.

Hence our simulations find the truncated circle Dirac operator, which we know to be the correct solution.
This is a good test for the formalism, and encourages us to move on from the simple circle to the more complicated sphere.

\subsection{$S^2$ simulations}

The version of equation \eqref{eq:CCM} corresponding to the sphere $S^2$ is
\begin{align}\label{eq:2dCCM}
  \constraint(Y_\Lambda,D_\Lambda,\gamma_\Lambda) = \langle Y_\Lambda [ D_\Lambda,Y_\Lambda] [D_\Lambda,Y_\Lambda] \rangle -\gamma_\Lambda.
\end{align}
Here, $Y = \begin{pmatrix} z & x - i y \\ x + i y & -z \end{pmatrix}$, with $x,y,z$ the standard coordinates on $\R^3$, viewed as functions on $S^2$ through its standard embedding. That is, $Y-1$ is twice the Bott projector.
The angular brackets denote the $B(P_\Lambda H)$-valued trace on $M_2(B(P_\Lambda))$ and $\gamma_\Lambda$ is the truncation of the usual grading on $L^2(S^2,S)$.
See the Appendix~\ref{appendix:representation} for the representation used in the numerical simulations.

For the sphere the Dirac operator has a few symmetries that the truncated operator should satisfy for the truncated operator to still interact correctly with the truncated chirality and real structure.
This leads us to consider different parametrizations for the operator.

\subsubsection{Parametrizing the Dirac operator}
In order to cancel the symmetry $D \mapsto D + c I$ of \eqref{eq:2dCCM} and to enforce symmetry of the spectrum of $D_\Lambda$, we have tested two different additional constraints.
The first, stronger constraint is that $D_\Lambda$ correspond to the (truncation of the) same K-cycle as $D_{S^2}$.
The second, strictly weaker constraint is that $D_\Lambda$ anticommute with $\gamma_\Lambda$, which is necessary for $D_\Lambda$ to possibly correspond to part of an even spectral triple $(C^\infty(S^2), H, D, \gamma)$.
These constraints lead to the parametrizations
$$
D_\Lambda = \begin{pmatrix}-P & 0 \\ 0 & P \end{pmatrix}
\qquad \text{or} \qquad
D_\Lambda = \begin{pmatrix}R & S \\ -S & -R \end{pmatrix},
$$
respectively, where $P$ is positive (ensuring that $D_\Lambda = D_\Lambda^*$ and $\operatorname{sign}(D) = \operatorname{sign}(D_{S^2})$) and $R, iS$ selfadjoint (ensuring that $D_\Lambda = D_\Lambda^*$ and $D_\Lambda \gamma_\Lambda = - \gamma_\Lambda D_\Lambda$).

The former parametrization is faster than the latter since both eigendecompositions of $P$ and the search for optimal $D_\Lambda$ occur in a vector space of half the dimension.
The geometries parametrized through $P$ are strictly a subclass of those parametrized through $R,S$, hence we know that solutions arising in the first ensemble also exist in the second.
Our simulations however show that to find the same optimal solutions in the $R,S$ parametrization requires longer runtimes and much lower temperatures.
This is because the larger configuration space takes longer to explore and lowers the relative fraction of the most optimal solutions.
We have tested that the $R,S$ simulations do not allow for additional, more optimal solutions than the $P$ parametrization, hence the results shown will all use the $P$ parametrization.

\subsubsection{Results}
\begin{figure}
  \centering
  \includegraphics[width=\maxfigwidth,height=\maxfigheight,keepaspectratio]{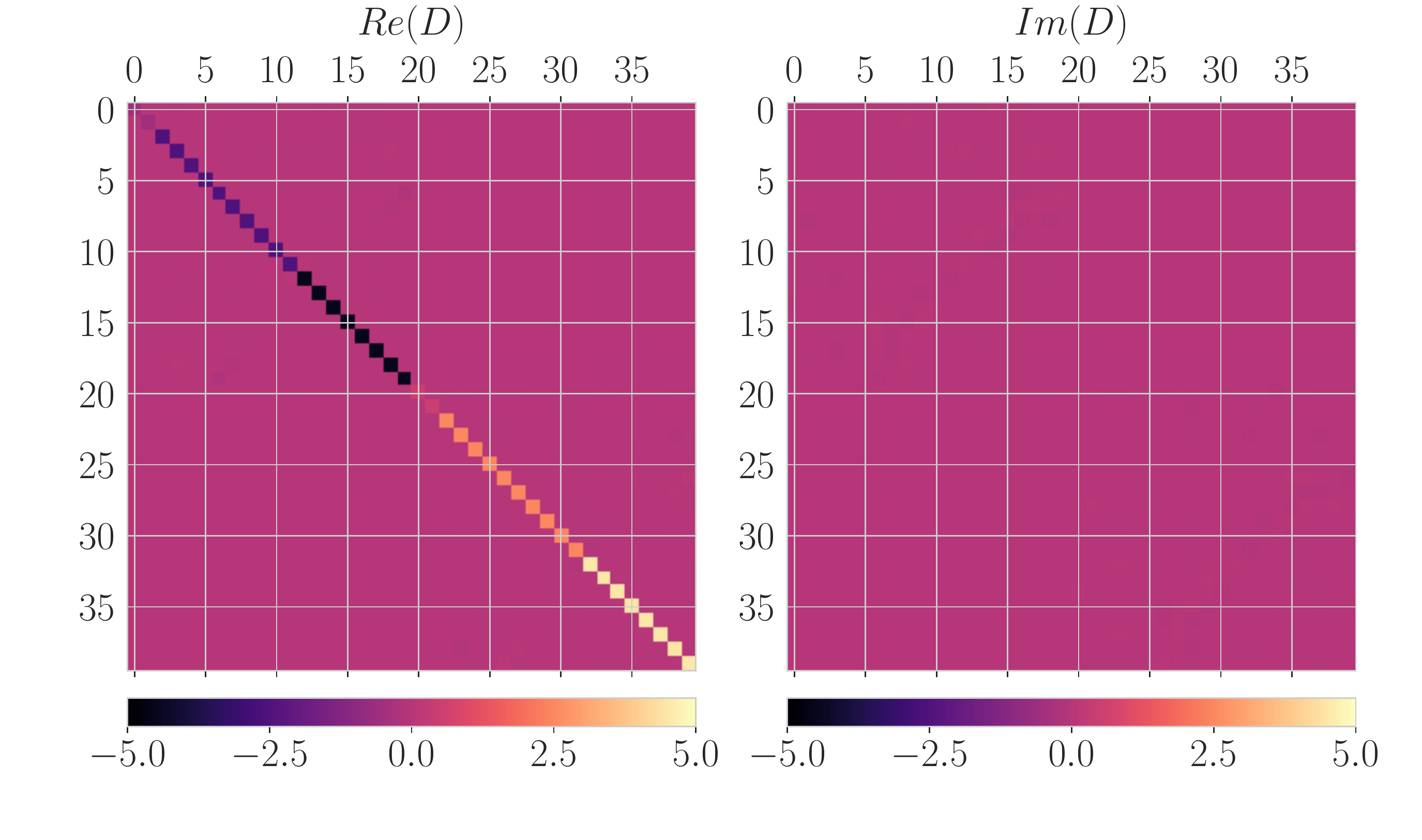}
\caption{The average Dirac operator is almost entirely real, and completely diagonal.}
    \label{fig:DPRS}
\end{figure}

To visualize the results of our simulations we will again look both at averages over roughly $150$ measurements near the minimum as well as at the actual numerical minimum of equation~\eqref{eq:2dCCM} that was encountered.
If we look at the operators as heatmaps, see Figure~\ref{fig:DPRS}, we see that the average Dirac operator in the $-P \oplus P$ parametrization commutes (up to numerical error) with $\DS$.

\begin{figure}
\includegraphics[width=\maxfigwidth,height=\maxfigheight,keepaspectratio]{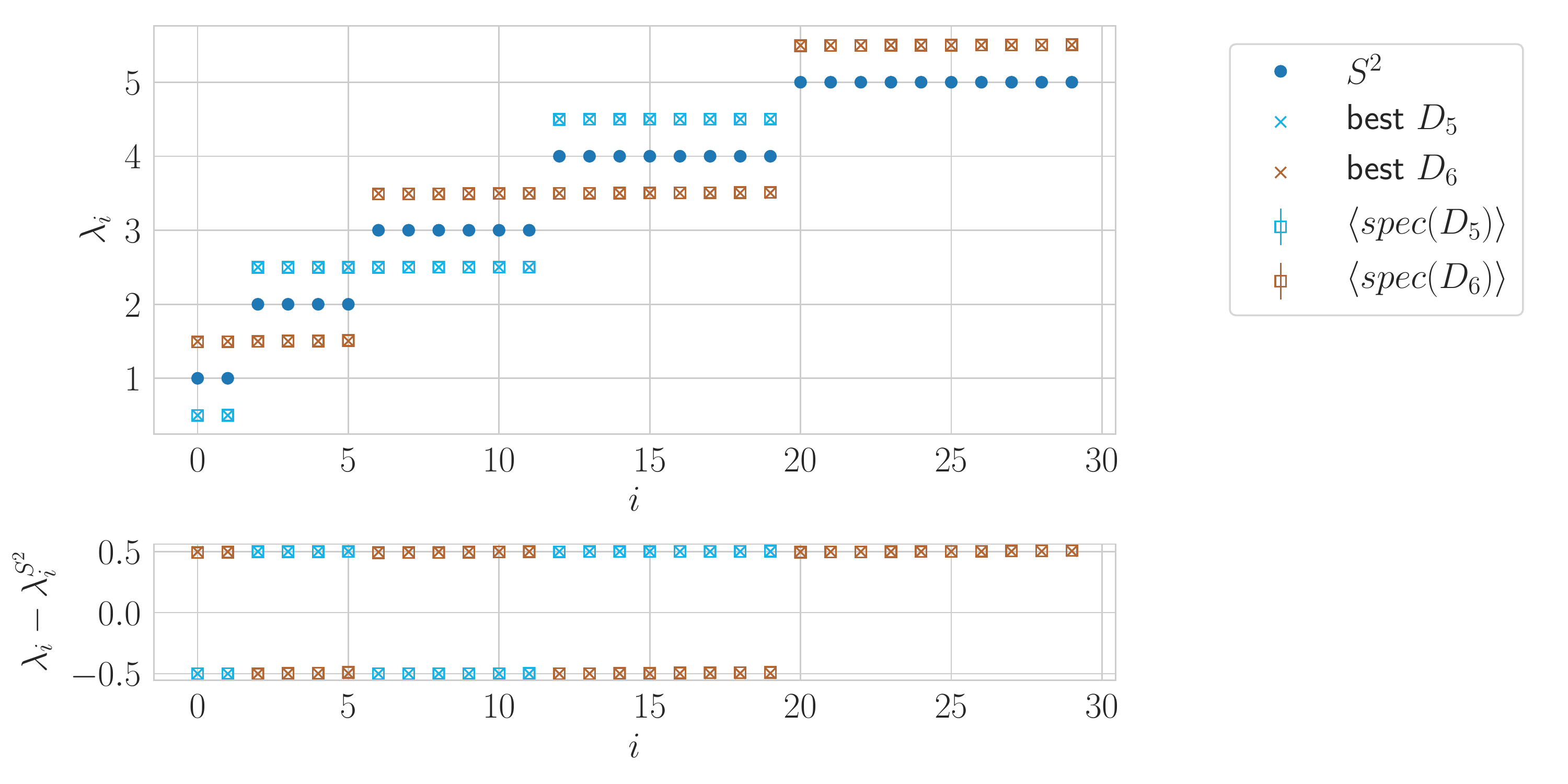}
\caption{Comparing the average eigenvalues, and the best case eigenvalues of the simulations with those of the sphere. We can see that the results differ considerably between odd and even $\Lambda$, but that neither agrees with the sphere.}
\label{fig:EV_sphere}
\end{figure}

This simple structure of the simulated Dirac operators implies they are well described, quantitatively, by their spectrum.
In Figure~\ref{fig:EV_sphere}, we compare the measured eigenvalues with those of the sphere.
The Figure shows results for spectral cutoffs of $\Lambda=5, 6$, which showcases a clear difference between odd and even cutoffs.

The simulated Dirac operators are (up to numerical error) diagonal in an eigenbasis of $\DS$, but the simulated eigenvalues are shifted up or down by roughly $\frac12$.
The direction of the shift appears dependent on the parity of the eigenvalue and of the cutoff $\Lambda$.
That is to say, it seems we are be dealing with a bounded perturbation of $\DS$ with particularly simple structure.

In particular, the localized zeta function asymptotics (which measure at least volume and dimension) must agree for this perturbation and the sphere. When we have identified the numerical solutions analytically, in Section~\ref{sec:analytic} below, we will show in Figure~\ref{fig:spectral_asymptotics} how this fact is reflected by the finite parts of the spectrum obtained.

\subsubsection{Results for the Heisenberg equation}
\begin{figure}
  \centering
  \subfloat[The exact sphere]{\includegraphics[width=\maxfigwidth,height=\maxfigheight,keepaspectratio]{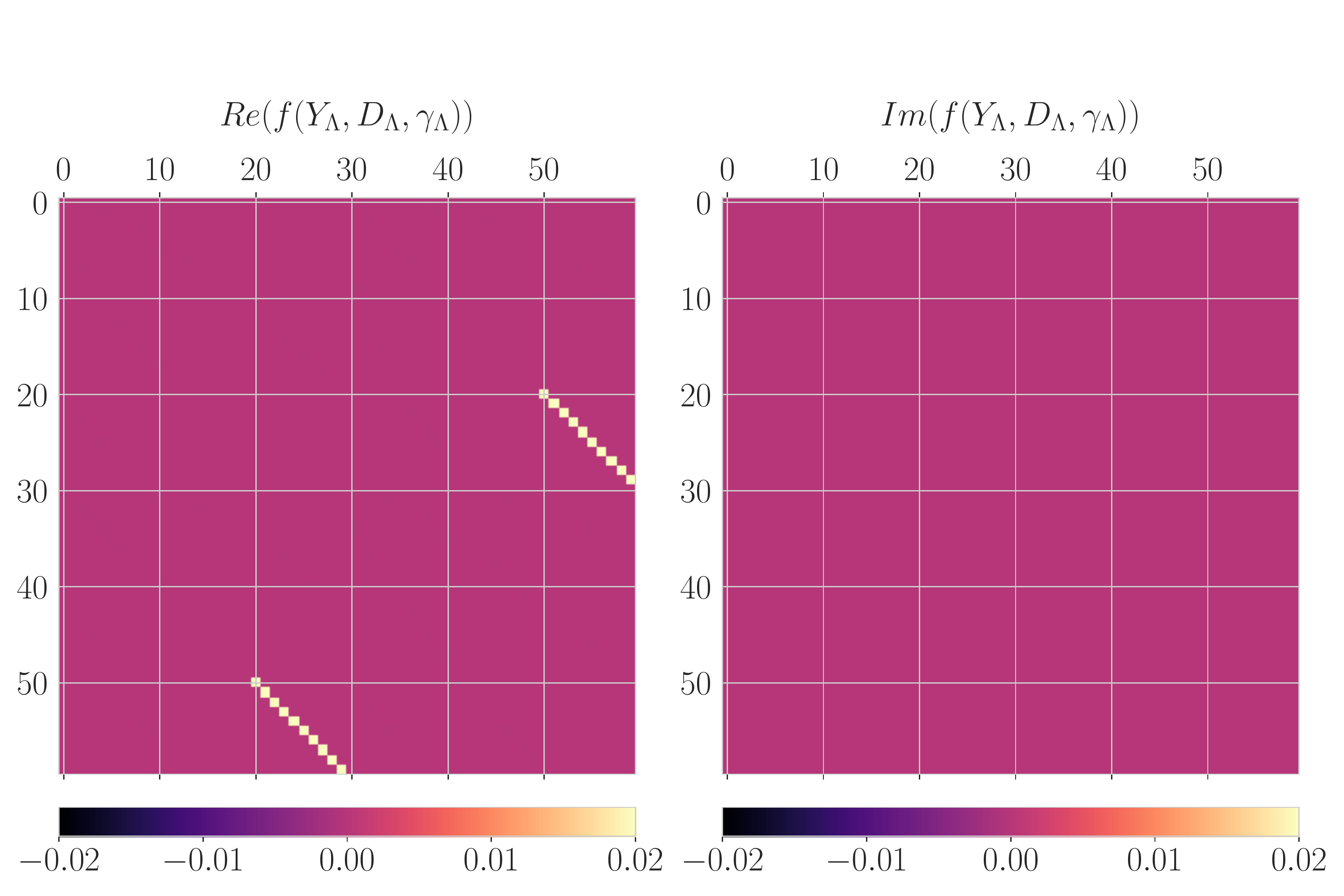} }

  \subfloat[Average operator]{  \includegraphics[width=\maxfigwidth,height=\maxfigheight,keepaspectratio]{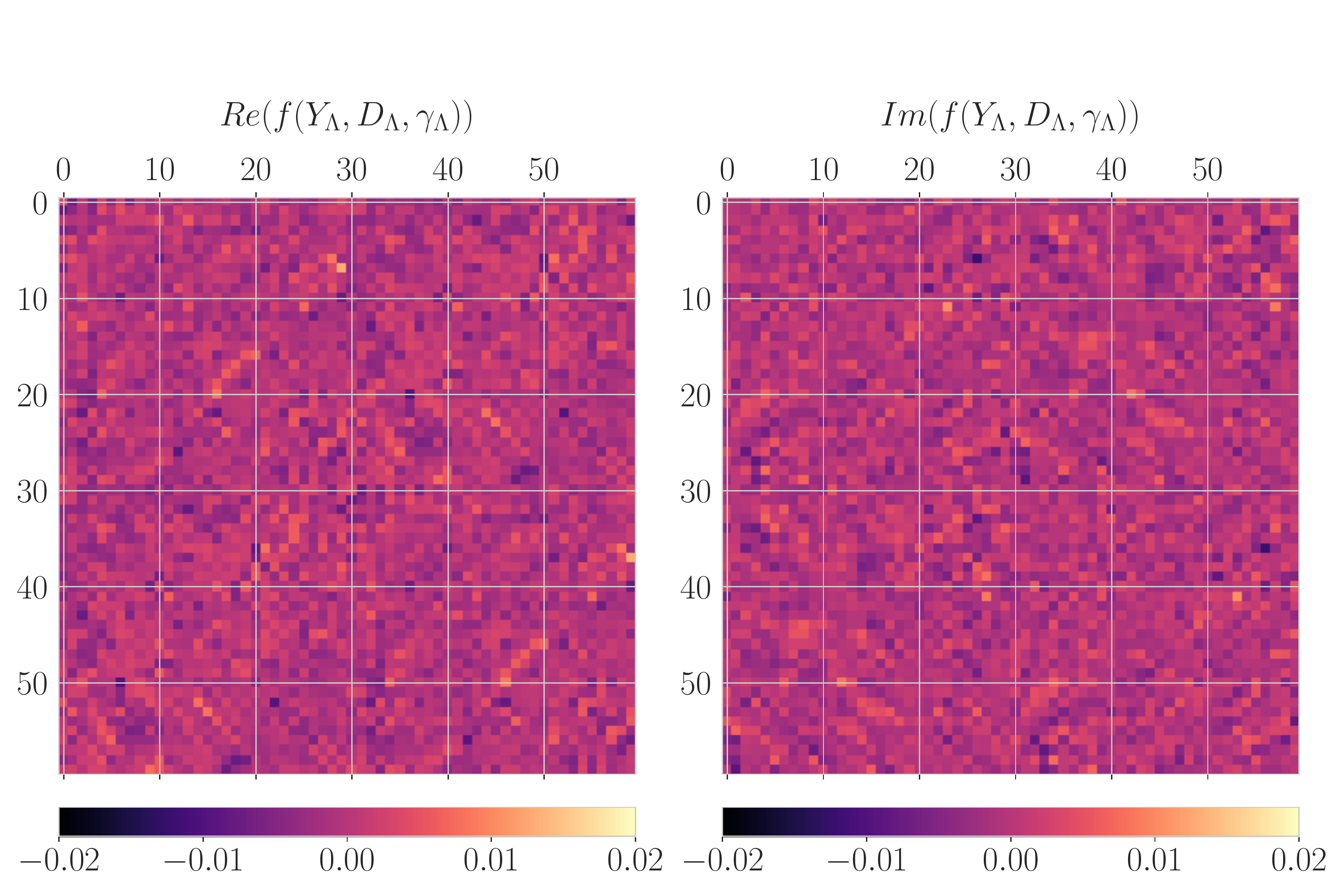} }

  \caption{Heatmap plot of the Heisenberg relation for the operator parametrized through $P$ and the sphere for $\Lambda=6$. The uppermost plot shows the finite size defects in the sphere, while the lower plots show the defect generated by an averaged Dirac operator.}
  \label{fig:CCM_sphere}
\end{figure}

The operators in Figure \ref{fig:DPRS} arise from minimization of the Heisenberg constraint $\norm{\constraint(Y_\Lambda,\gamma_\Lambda,D)}^2_2$, so it is interesting to see whether patterns arise in the corresponding matrix entries of $\constraint(Y_\Lambda,\gamma_\Lambda,D)$; we show these in Figure \ref{fig:CCM_sphere}.
Clearly, the simulations come close to fully letting $\constraint(Y_\Lambda,\gamma_\Lambda,D)$ vanish.

For the operator $D_{S^2,\Lambda}$, however, the defect $\constraint(Y_\Lambda,D_{S^2,\Lambda},\gamma_\Lambda)$ does not vanish and equals
\begin{equation}
  \constraint(Y_\Lambda,D_{S^2,\Lambda},\gamma_\Lambda) = - \frac{(1 + \lambda)(1 + 4 \lambda)}{2(1+2\lambda)^2} (E_\lambda + E_{- \lambda}) \gamma
\end{equation}
where $E_\lambda$ projects onto the eigenspace corresponding to $\lambda = \max \{\lambda' \in \sigma(D) \mid |\lambda'| \leq \Lambda \}$; this is of norm $\sim \frac12$ and of divergent $( O(\Lambda^{1/p}))$ $p$-Schatten norm for $p < \infty$ as $\Lambda \to \infty$.

\begin{figure}
  \includegraphics[width=\maxfigwidth,height=\maxfigheight,keepaspectratio]{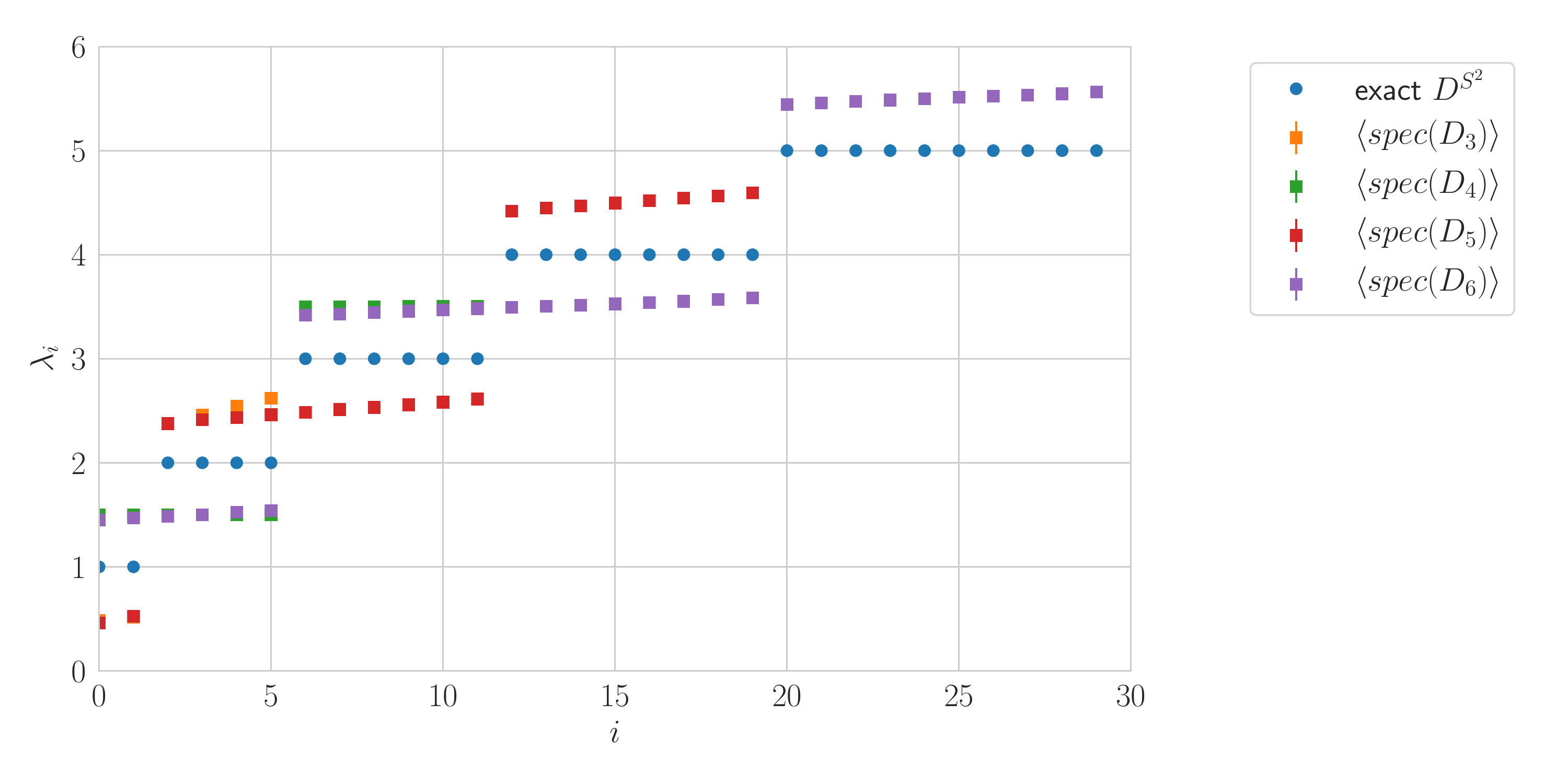}
  \caption{Average eigenvalues for the $4$ smallest truncations of the sphere.}
  \label{fig:smallTrunc}
\end{figure}

For each $\Lambda$ considered, we found a $D_\Lambda$ with $\| \constraint(Y_\Lambda, \gamma_\Lambda, D_\Lambda) \|_2 \approx 0$ and in particular $\|\constraint(Y_\Lambda, \gamma_\Lambda, D)\|_2 \ll \| \constraint(Y_\Lambda, \gamma_\Lambda, D_{S^2,\Lambda})\|_2$.
Additionally these optimal $D_\Lambda$ seem to be quite simple and symmetric, and shows a remarkable consistency across different sizes, as shown in Figure~\ref{fig:smallTrunc}.
Since the matrix size (the rank of $P_\Lambda$) grows as $O(\Lambda^2)$ it is hard to obtain reliable results for larger $\Lambda$, however the results we obtained suggest that there might be a similar type of solution for all sizes, i.e. a compatible chain of finite size Dirac operators that might arise as $P_\Lambda D' P_\Lambda$ for some $D'$ that solves \eqref{eq:CCM} exactly.
It is thus useful to supplement the numerical results with some analytic explorations.

\section{An alternative analytic solution to the Heisenberg relation}
\label{sec:analytic}
The simulations above suggest that, for finite $\Lambda$, there might be a class of operators $D \in B(P_\Lambda H)$ that lead to lower values of the constraint $\norm{\constraint(Y_\Lambda,\gamma_\Lambda,D)}_2$ than the truncations of $\DS$ do.
Since the $D$ that show up commute with $\DS$ and seem to be compatible across alternating choices of $\Lambda$ (see Figure~\ref{fig:smallTrunc}), we are led to look analytically for a corresponding general solution of $\constraint(Y,\gamma,D) = 0$ inside the commutant of $\DS$.

Let us denote by $\fdspace$ the space of selfadjoint operators with discrete spectrum that commute with $D_{S^2}$ and anticommute with $\gamma$, that is, those of the form $f(D)$ for some antisymmetric $f \in C(\R,\R)$.
Is there an analytic solution $D \in \fdspace$ to equation \eqref{eq:2dCCM}?

The Appendix~\ref{appendix:representation} exhibits the coefficients of the representation of $Y, \gamma, D$ on $H$ in the basis chosen for the simulations.
Since the generators $Y_i$ are laddering, i.e. band, matrices in this basis, the resulting version of equation~\eqref{eq:2dCCM} is easy to solve analytically.

It leads to the following recursion for the sequence $\mu_l$ of positive eigenvalues of $D$, labeled by the spinor momenta $l = \frac12, \frac32, \dotsc$,
\begin{align*}
  \mu_l^2 - 2 a_l \mu_{l} \mu_{l-1} +  2 b_l \mu_{l+1} \mu_l =
  a_l \mu_{l-1}^2  + b_l \mu_{l+1}^2 + 16 l^2 (1 + l)^2
\end{align*}
where $a_l = (1+l)^2(2l-1)$, $b_l = l^2 (3 + 2 l)$
The corresponding recursion equation, with $\mu_{-1/2} = 0$, has the unique one-parameter solution
\[
 \mu_l = \left(l +\frac12\right) + c \sin(\pi l).
\]
That is to say, the unique one-parameter solution $\{D_c \mid c \in \R\} \subset \fdspace$ to equation \eqref{eq:2dCCM} is
\[
  D_c = D_{S^2} + c B,
\]
where the bounded and selfadjoint operator $B$ equals $\operatorname{sign}(\DS) \cos(\pi \DS)$.

Looking at this we see that it agrees with the Dirac operator we found in our simulations using the parametrization with $P$.
In particular for $c=\pm 1/2$ this agrees with the simulations with an even/ odd maximal eigenvalue, as shown in Figures \ref{fig:EV_sphere} and \ref{fig:smallTrunc}.

\subsection{Spectral triple axioms}

For nonzero $c$, the full operator $D_{S^2} + cB$ does not satisfy the first-order axiom (condition 2 in the reconstruction theorem of \cite{Connes:SpectralCharacterizationManifolds}) because $[[B,Y_i],Y_j]$ is not zero for all $i,j$; $B$, although pseudodifferential of order zero, is not an endomorphism of the spinor bundle.
The defect $[[B,Y_i],Y_j]$, however, is compact (it is in fact in $L^{(1,\infty)}(H)$).
As we will see in the next subsection the boundary effects caused by the truncation to finite matrix sizes mask this difference and lead to violations of the first order axiom for $D_{S^2}$ alone that are of the same order of magnitude as the violation for $B$.

\subsection{Boundary defects}
As mentioned in the introduction, replacing a solution $Y, \gamma, D$ of the one-sided higher Heisenberg equation \eqref{eq:CCM} by $Y_\Lambda, \gamma_\Lambda, D_\Lambda$ leads to a nontrivial defect $\constraint(Y_\Lambda,\gamma_\Lambda,D) = \langle Y_\Lambda [D_\Lambda,Y_\Lambda]^n \rangle - k n! \gamma_\Lambda$.

\begin{figure}
  \centering
  \includegraphics[width=\maxfigwidth,height=\maxfigheight,keepaspectratio]{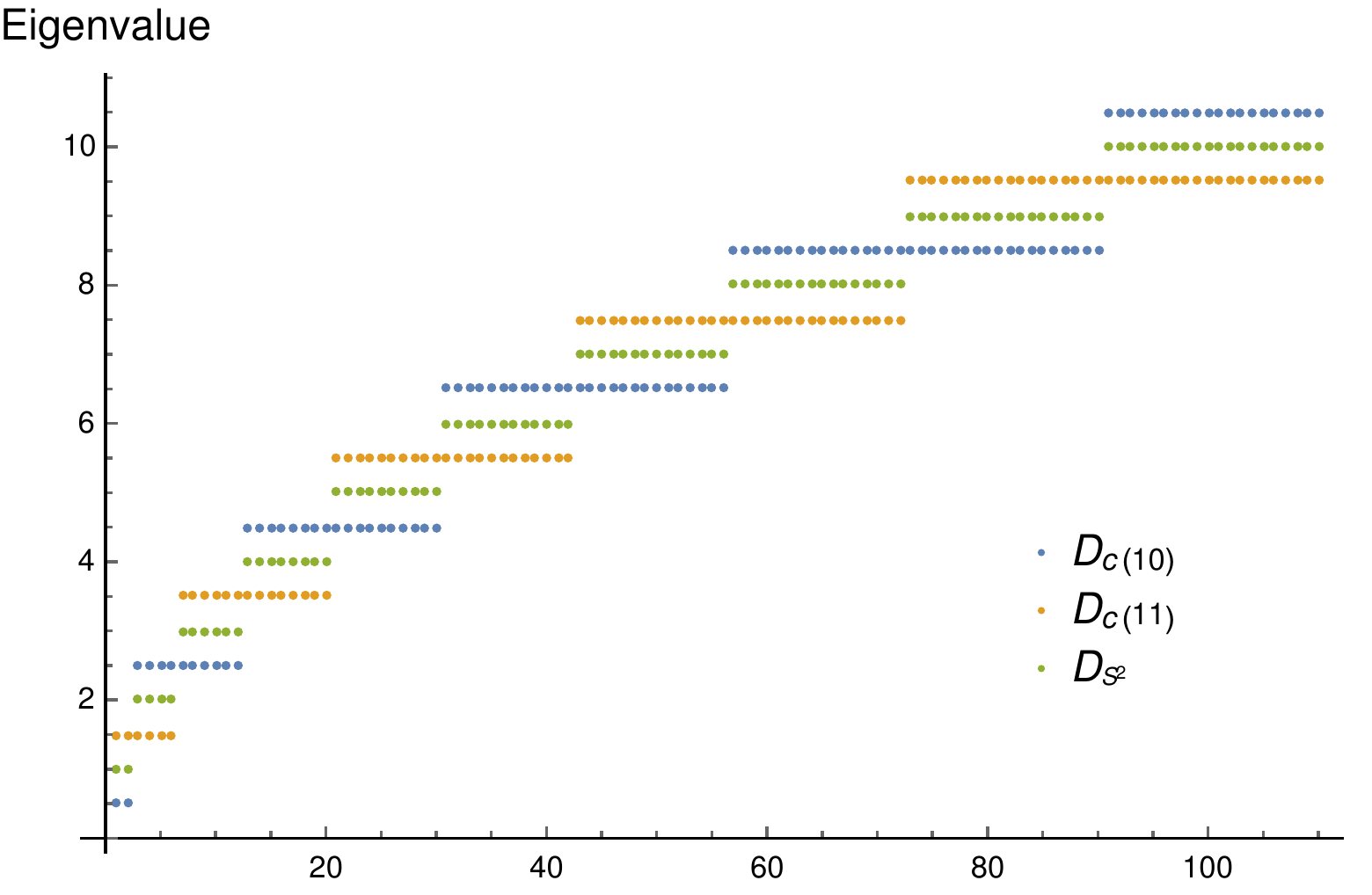}
  \caption{Spectra of $\DS$ and $\DB$ for even/odd $\lambda_{\mathrm{max}}$.}
  \label{fig:exactspectra}
\end{figure}

For operators in $\fdspace$ this introduces an additional term $\constraint(Y_\Lambda,\gamma_\Lambda,D)$ as compared to $\constraint(Y,\gamma,D)$. This term is a multiple of the $\gamma$ operator projected onto the highest eigenspace of $|D|$, where the coefficient equals
\[
  c_l \mu_l^2  + \frac{(1-2l)}{16 l^2} \mu_{l-1}(\mu_{l-1} + 2 \mu_l) - 1,
\]
with $c_l = \frac{1 + 9 l^2 + 6 l^3}{16 l^2(l+1)^2}$.
In terms of the parameter $c$, above, this means that additionally to solving $\constraint(Y,\gamma,D_c) = 0$ we can solve the finite-cutoff equation $\constraint(Y_\Lambda,\gamma_\Lambda,D_c) = 0$ (uniquely) by $c = s(\Lambda) / 2$, where the sign $s(\Lambda)$ equals the parity $\cos(\pi \lambda_{\mathrm{max}})$ of the highest eigenvalue $\lambda_{\mathrm{max}}$ of $|D_{S^2}|$ below $\Lambda$ (so that the corresponding eigenvalue of $cB$ is $+\frac12$): see Figure~\ref{fig:exactspectra}.

The finite-rank operators $D_{c,\Lambda}$, for any $c \in \R$, never satisfy the first-order condition that $[[D_{\Lambda},Y_{\Lambda}],Y_{\Lambda}]$ should vanish.
For $D = D_{S^2}$, for which the defect vanishes in the strong limit $\Lambda \to \infty$, there is a boundary defect of asymptotically constant norm (that is, $\| [[D_{S^2,\Lambda},Y_{\Lambda}],Y_{\Lambda}\| \sim 1$) and of unbounded trace norm, (that is, $\|[[D_{S^2,\Lambda},Y_{\Lambda}],Y_{\Lambda}\|_1 = O(\Lambda)$) as $\Lambda \to \infty $.

As mentioned above, the defect $[[B_{\Lambda},Y_{\Lambda}],Y_{\Lambda}]$ does not vanish in the strong limit $\Lambda \to \infty$.
However, precisely when $c = s(\Lambda)/2$ as above, the highest-order terms of $[[c B_{\Lambda},Y_{\Lambda}],Y_{\Lambda}]$ and $[[D_{S^2,\Lambda},Y_{\Lambda}],Y_{\Lambda}]$ cancel each other.
As a result, the defect $[[D_{c,\Lambda},Y_{\Lambda}],Y_{\Lambda}]$ is of norm $O(\Lambda^{-1})$ and trace norm $O(1)$.
In this sense it is hard to computationally detect the fact that $D_{S^2,\Lambda}$ comes from a spectral triple while (for nonzero $c$) $D_{c,\Lambda}$ does not.

\subsection{Visualisation: a locally isometric graph embedding}
\label{sec:embedding}
The operator $\DB$ seems, at least on $P_\Lambda H$ for finite $\Lambda$, to come closer to satisfying the higher Heisenberg equation \eqref{eq:CCM} than the original solution $\DS$ does, and neither its spectral asymptotics nor the first-order equation allow us to discern at the finite level that it does not form a commutative spectral triple with $C^\infty(S^2)$ and $L^2(S^2,S)$.
This suggests to pretend it \emph{does} arise from a spin geometry and to compare at least the resulting metric on $S^2$ to the standard one.

\begin{figure}
  \centering
  \subfloat[Dimension estimate]{
  \includegraphics[width=\maxfigwidth,height=\maxfigheight,keepaspectratio]{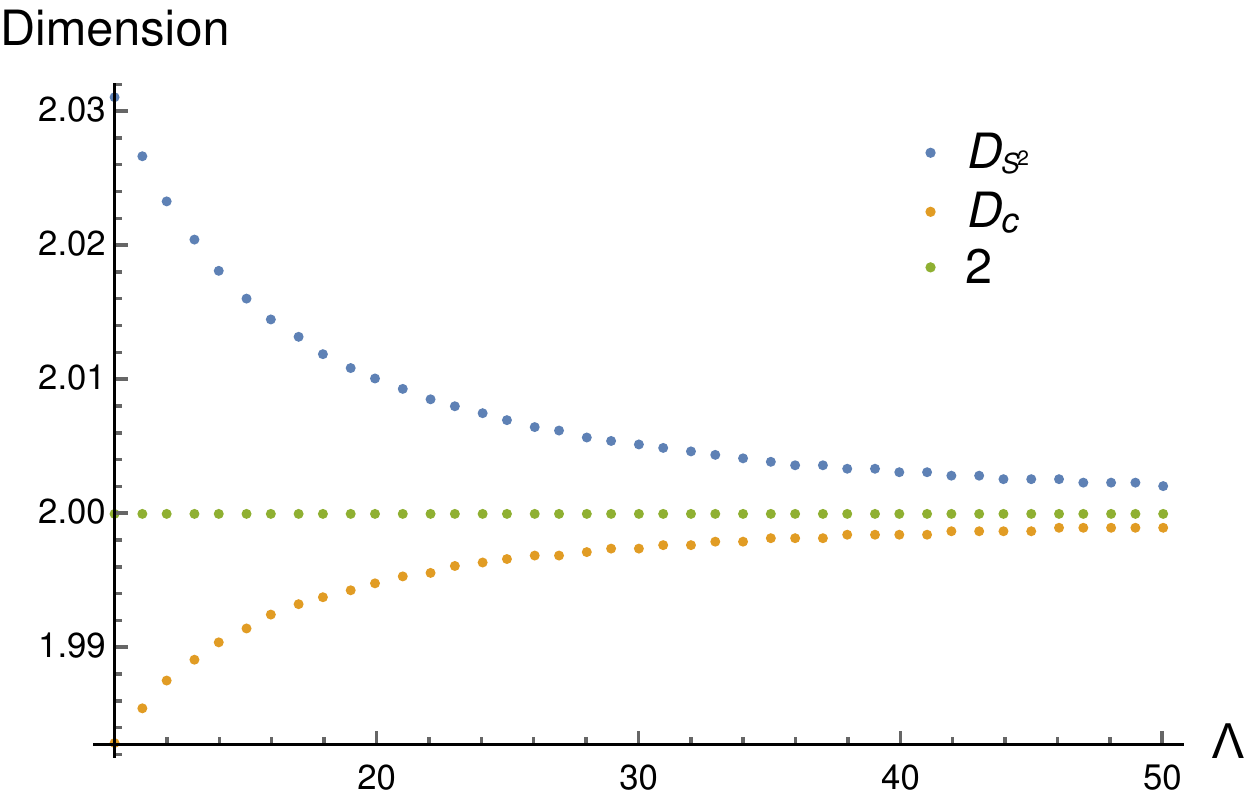}}
\newline
\subfloat[Volume estimate]{
  \includegraphics[width=\maxfigwidth,height=\maxfigheight,keepaspectratio]{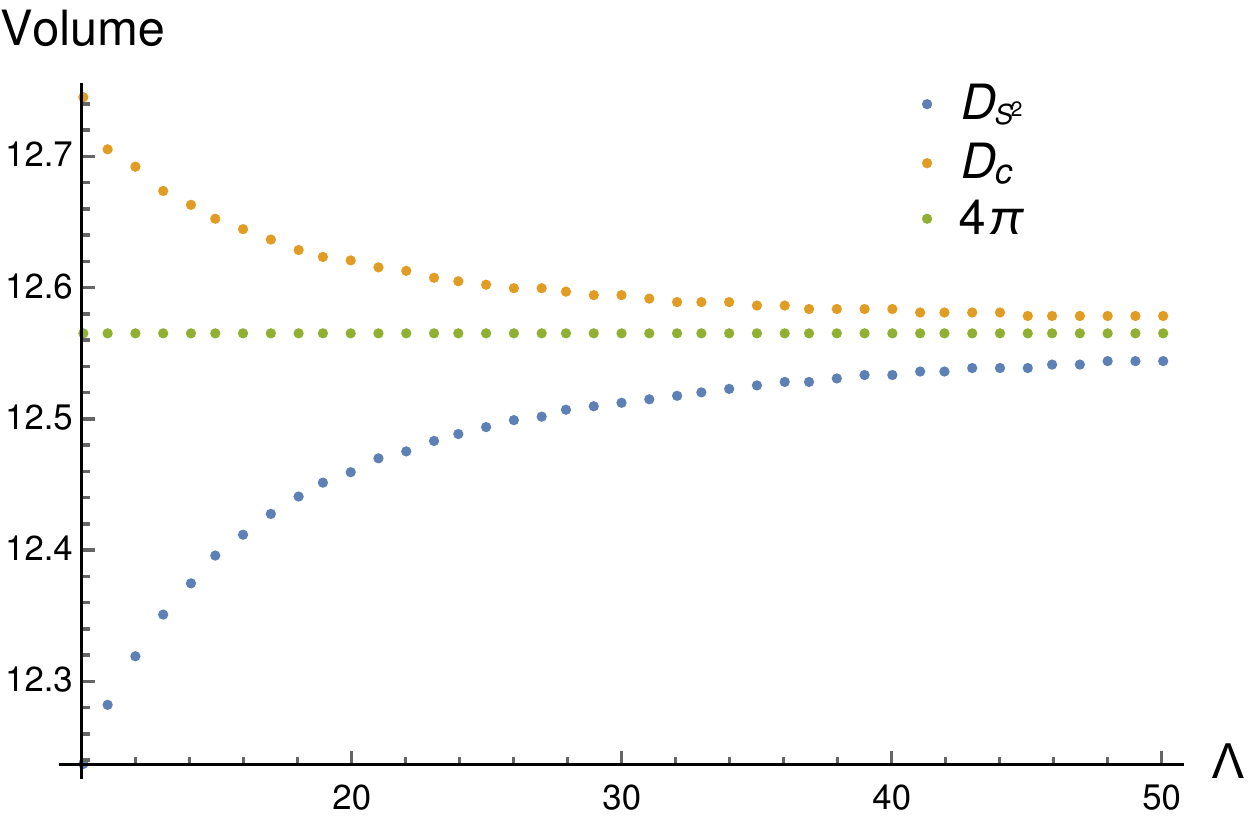}}
\caption{Finite-rank estimates of the spectral asymptotics of $\DS$ and $\DB$}
\label{fig:spectral_asymptotics}
\end{figure}

First of all, since the difference $B$ is bounded, the Weyl asymptotics agree in the sense that the first zeta residues must be equal in both value and argument.
This is already detectable at the truncated level, e.g. using the finite-rank zeta approximations from \cite{Stern_2018}: see Figure \ref{fig:spectral_asymptotics}.
One interesting feature of these figures is that the dimension and volume estimators converge faster for the $\DB$ operator than for the truncated sphere.

The asymptotics corresponding to total scalar curvature, however, are completely different for $\DB$ (the corresponding residue is not $\frac{2 \cdot 4 \pi}{6 \cdot 4 \pi}$ but rather $\frac{- 4 \pi}{6 \cdot 4 \pi}$) because it is the $O(t^{-n/2+1})$ term in the asymptotics of $\tr e^{- t D^2}$ and is therefore highly sensitive to bounded shifts when the dimension equals $2$.

The companion paper~\cite{GlaserSternDistance19} develops a method to associate a finite metric space to `operator system spectral triples' $(P_\Lambda C^\infty(M) P_\Lambda, P_\Lambda H, P_\Lambda D)$.
The method, briefly, is as follows.

\begin{figure}
  \centering  \subfloat[$\DS$]{\includegraphics[width=\maxfigwidth,height=\maxfigheight,keepaspectratio]{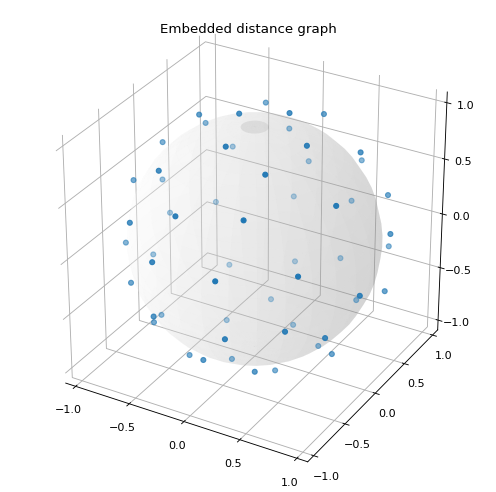}}
  \newline
\subfloat[$\DB$]{\includegraphics[width=\maxfigwidth,height=\maxfigheight,keepaspectratio]{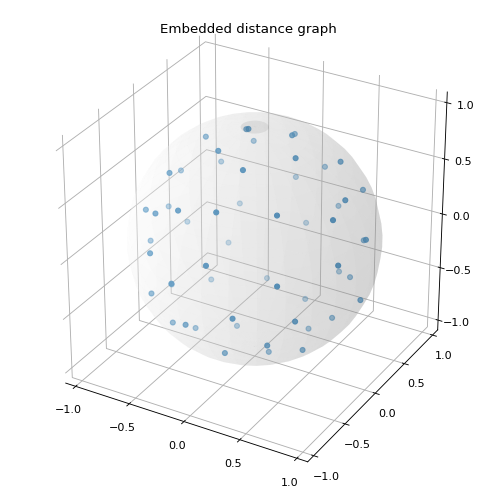}}
\caption{Locally almost-isometric embeddings corresponding to $\DS$ and $\DB$, with shaded $S^2$ for reference}
\label{fig:visualisation}
\end{figure}

\begin{itemize}
  \item The embedding $Y$ is used to define the \emph{dispersion} $\delta(v) \eqdef \sum_{i} \langle v, Y_i^2  v \rangle - \langle v, Y_i v \rangle^2$ of a vector $v \in H$, which measures the degree to which the corresponding vector state is localized. In the commutative case, this corresponds to the statistical variance of the position variable $Y$ under the measure induced by $v$.
  \item One iteratively constructs a reasonably dense (finite) set of localized states by minimizing the dispersion, combined with an electrostatic repulsion to avoid repetition. Up to the distortion induced by imperfect localization, this results in the commutative case in generating a set of roughly equidistributed points on the underlying manifold.
    \item The Connes distance formula (\ref{eq:distance}) is used to calculate the distance between the generated states, in order to obtain a metric graph. In the commutative case, those distances correspond to the Kantorovich-Wasserstein distance between the measures induced by the localized states, which reduces to the geodesic distance in the limit of perfect localization.
   \item The SMACOF algorithm is utilized to embed the obtained metric graph in $\R^n$ in an asymptotically locally isometric way. This means that, asymptotically as $\Lambda \to \infty$, the embedding is pressured to be Riemannian.
    \end{itemize}
\noindent For $\DS$ and $\DB$, this procedure yields the images displayed in Figure \ref{fig:visualisation}.

\section{Conclusions}
\label{sec:conclusion}
In this article we explored the behaviour of the truncated one-sided higher Heisenberg relation in dimensions 1 and 2.
In the one-dimensional case the simulations yielded the expected result, showing that the truncation of the Dirac operator on the circle is closest to solving the corresponding truncated relation.
The two-dimensional version of the truncated Heisenberg relation, however, lead to a new minimum that differs from (but commutes with) the truncated Dirac operator on the sphere.
We found analytically that this numerical minimum corresponds to the truncation at $c = \pm \frac12$ of a new one-parameter family $D_c = \DS + c B$ of exact solutions to the non-truncated higher Heisenberg equation.
While these bounded perturbations $D_c$ of $\DS$ satisfy most conditions of the reconstruction theorem, they fail to satisfy the first-order condition. Unlike many other geometric properties, however, this defect turns out to not be detectable at the truncated level.

An interesting comparison here is the case of the four-dimensional version of the higher Heisenberg relation.
That relation is solved not only by the four-sphere, but also by an additional, genuinely non-commutative, spectral triple, the Connes-Landi sphere~\cite{Connes2002}.
This similarity invites the question whether the Heisenberg relation might invite more freedom the larger the dimension becomes.

There are many interesting extensions of this work waiting to be explored.
In particular the Heisenberg relation needs to be understood in more detail.
It is unclear how its one-sided version behaves in higher dimensions and, just as importantly, when more freedom is allowed for the parameter $Y$.
Our results, seen in context with the Connes-Landi sphere, suggest that more conditions are required to ensure that we deal with truncations of genuine Dirac spectral triples.
In addition, it would be interesting to explore the two-sided equation of the Heisenberg relation. In that context, allowing the embedding maps $Y$ to vary as well as the Dirac operator enlarges the resulting ensemble to contain \emph{all} spin manifolds of the dimensions considered.
With additional conditions, this would be a solid basis for a spectral version of random geometry, which could be compared to and begin a dialogue with results in quantum gravity, such as those of dynamical triangulations~\cite{Loll_1998} and spinfoams~\cite{Perez_2013}.

\section*{Acknowledgements}
We would like to thank Walter van Suijlekom for extensive discussions starting this project.
LG has been funded through grant number M 2577 through the Lise Meitner-Programm of the FWF Austria.
ABS has been funded through FOM Vrij Programma No. 150.

\appendix
\section{Representation of $Y, \gamma, \DS$}
\label{appendix:representation}

Let $S$ be the standard spinor bundle over $S^2$, with Dirac operator $\DS$, and let $x, y, z$ be the standard coordinate functions on $S^2 \subset \R^3$.
Then, the spectral triple $(C^\infty(S^2), L^2(S^2; S), \DS)$ can be represented as follows.
Let $\{ \ket{l,m}_\pm \mid l \in \Z_{\geq 0} + \frac12, -l \leq m \leq l \}$ be an orthonormal basis of the Hilbert space $H$.
Then, we represent the generators $a = 2 (x - i y)$ and $b = z$ of the algebra $C^\infty(S^2)$, the grading $\gamma$ of $S$ and the Dirac operator $\DS$ as follows:

\begin{align*}
  a \ket{l,m}_\pm = & - \frac{\sqrt{(l+m+1)(l-m)}}{2l(l+1)} \ket{l,m+1}_\mp \\
                    & + \frac{\sqrt{(l+m+1)(l+m+2)}}{2(l+1)} \ket{l+1,m+1}_\pm \\
                    & - \frac{\sqrt{(l-m)(l-m-1)}}{2l} \ket{l-1,m}_\pm, \\
  b \ket{l,m}_\pm = & \frac{m}{2 l(l+1)} \ket{l,m}_\mp \\
                    & + \frac{\sqrt{(l-m+1)(l+m+1)}}{2(l+1)} \ket{l+1,m}_\pm \\
                    & + \frac{\sqrt{(l-m)(l+m)}}{2l} \ket{l-1,m}_\pm,  \\
  \gamma \ket{l,m}_\pm = & \ket{l,m}_\mp, \\
  \DS \ket{l,m}_\pm = & \pm (l + \frac12) \ket{l,m}_\pm.
 \end{align*}
 This representation was chosen to align well with that of \cite{dabrowski2005spectral}.
 We then write the matrix $Y$ as $\begin{pmatrix} b & a \\ a^* & -b \end{pmatrix}$.

 \bibliographystyle{plain}
 \bibliography{bib}

\begin{thebibliography}{10}

\bibitem{BALACHANDRAN2002184}
A.P. Balachandran, Brian~P. Dolan, J.~Lee, X.~Martin, and Denjoe O'Connor.
\newblock Fuzzy complex projective spaces and their star-products.
\newblock {\em Journal of Geometry and Physics}, 43(2):184 -- 204, 2002.

\bibitem{Barrett:2015naa}
John~W. Barrett.
\newblock {Matrix geometries and fuzzy spaces as finite spectral triples}.
\newblock {\em J. Math. Phys.}, 56(8):082301, 2015.

\bibitem{Barrett:2019aig}
John~W. Barrett, Paul Druce, and Lisa Glaser.
\newblock {Spectral estimators for finite non-commutative geometries}.
\newblock {\em J. Phys.}, A52(27):275203, 2019.

\bibitem{Barrett_Glaser_2016}
John~W. Barrett and Lisa Glaser.
\newblock Monte carlo simulations of random non-commutative geometries.
\newblock {\em J.Phys.}, A49:245001, May 2016.

\bibitem{cacic2011moduli}
Branimir {\'C}a{\'c}i{\'c}.
\newblock Moduli spaces of dirac operators for finite spectral triples.
\newblock In {\em Quantum Groups and Noncommutative Spaces}, pages 9--68.
  Springer, 2011.

\bibitem{Chamseddine:1996zu}
Ali~H. Chamseddine and Alain Connes.
\newblock The {Spectral} {Action} {Principle}.
\newblock {\em Communications in Mathematical Physics}, 186(3):731--750, July
  1997.
\newblock arXiv: hep-th/9606001.

\bibitem{chamseddine_spectral_1997}
Ali~H. Chamseddine and Alain Connes.
\newblock The {Spectral} {Action} {Principle}.
\newblock {\em Communications in Mathematical Physics}, 186(3):731--750, July
  1997.
\newblock arXiv: hep-th/9606001.

\bibitem{chamseddine2008standard}
Ali~H Chamseddine and Alain Connes.
\newblock Why the standard model.
\newblock {\em Journal of Geometry and Physics}, 58(1):38--47, 2008.

\bibitem{Chamseddine:2006ep}
Ali~H. Chamseddine, Alain Connes, and Matilde Marcolli.
\newblock {Gravity and the standard model with neutrino mixing}.
\newblock {\em Adv. Theor. Math. Phys.}, 11(6):991--1089, 2007.

\bibitem{C.C.M:GeometryQuantumBasics}
Ali~H. Chamseddine, Alain Connes, and Viatcheslav Mukhanov.
\newblock Geometry and the quantum: Basics.
\newblock {\em Journal of High Energy Physics}, 2014(12):98, 2014.

\bibitem{Connes_2000}
Alain Connes.
\newblock A short survey of noncommutative geometry.
\newblock {\em Journal of Mathematical Physics}, 41(6):3832, 2000.
\newblock arXiv: hep-th/0003006.

\bibitem{Connes:SpectralCharacterizationManifolds}
Alain Connes.
\newblock On the spectral characterization of manifolds.
\newblock {\em Journal of Noncommutative Geometry}, 7(1):1--82, 2013.

\bibitem{Connes2002}
Alain Connes and Giovanni Landi.
\newblock Noncommutative manifolds, the instanton algebra and isospectral
  deformations.
\newblock {\em Communications in Mathematical Physics}, 221(1):141--159, Jul
  2001.

\bibitem{ConnesSuij}
Alain Connes and Walter van Suijlekom.
\newblock {\em Work in preparation}.

\bibitem{dabrowski2005spectral}
Ludwik Dabrowski, Giovanni Landi, Mario Paschke, and Andrzej Sitarz.
\newblock The spectral geometry of the equatorial podle{\'s} sphere.
\newblock {\em Comptes Rendus Mathematique}, 340(11):819--822, 2005.

\bibitem{devicentePlacementThermodynamicSimulated2003}
Juan {de Vicente}, Juan Lanchares, and Rom{\'a}n Hermida.
\newblock Placement by thermodynamic simulated annealing.
\newblock {\em Physics Letters A}, 317(5):415--423, October 2003.

\bibitem{Dolan_2003}
Brian~P Dolan and Denjoe O'Connor.
\newblock A fuzzy three sphere and fuzzy tori.
\newblock {\em Journal of High Energy Physics}, 2003(10):060--060, oct 2003.

\bibitem{d2014spectral}
Francesco D’Andrea, Fedele Lizzi, and Pierre Martinetti.
\newblock Spectral geometry with a cut-off: topological and metric aspects.
\newblock {\em Journal of Geometry and Physics}, 82:18--45, 2014.

\bibitem{d2013metric}
Francesco D’Andrea, Fedele Lizzi, and Joseph~C V{\'a}rilly.
\newblock Metric properties of the fuzzy sphere.
\newblock {\em Letters in Mathematical Physics}, 103(2):183--205, 2013.

\bibitem{glaserScalingBehaviourRandom2017}
Lisa Glaser.
\newblock Scaling behaviour in random non-commutative geometries.
\newblock {\em Journal of Physics A: Mathematical and Theoretical},
  50(27):275201, 2017.

\bibitem{GlaserSternDistance19}
Lisa {Glaser} and Abel~B. {Stern}.
\newblock {Reconstructing manifolds from truncated spectral triples}.
\newblock {\em arXiv e-prints}, page arXiv:1912.09227, December 2019.

\bibitem{gracia2013elements}
Jos{\'e}~M Gracia-Bond{\'\i}a, Joseph~C V{\'a}rilly, and H{\'e}ctor Figueroa.
\newblock {\em Elements of noncommutative geometry}.
\newblock Springer Science \& Business Media, 2013.

\bibitem{Grosse1995}
H.~Grosse and P.~Pre{\v{s}}najder.
\newblock The dirac operator on the fuzzy sphere.
\newblock {\em Letters in Mathematical Physics}, 33(2):171--181, Feb 1995.

\bibitem{krajewski1998classification}
Thomas Krajewski.
\newblock Classification of finite spectral triples.
\newblock {\em Journal of Geometry and Physics}, 28(1-2):1--30, 1998.

\bibitem{Loll_1998}
Renate Loll.
\newblock Discrete approaches to quantum gravity in four dimensions.
\newblock {\em Living Reviews in Relativity}, 1(1):13, Dec 1998.

\bibitem{Perez_2013}
Alejandro Perez.
\newblock The spin-foam approach to quantum gravity.
\newblock {\em Living Reviews in Relativity}, 16(1):3, Feb 2013.

\bibitem{Stern_2018}
Abel~B. Stern.
\newblock Finite-rank approximations of spectral zeta residues.
\newblock {\em Letters in Mathematical Physics}, Jul 2018.

\end{thebibliography}

\end{document}